# The State of Computational Science in Fission and Fusion Energy

Andrea Morales Coto, Aditi Verma

*Abstract* – **The tools used to engineer something are just as important as the thing that is actually being engineered. In fact, in many cases, the tools can indeed determine what is engineerable. In fusion and fission[1] energy engineering, software has become the dominant tool for design. For that reason, in 2024, for the first time ever, we asked 103 computational scientists developing the codes used in fusion and fission energy about the problems they are attempting to solve with their codes, the tools available to them to solve them, and their end to end developer experience with said tools.**

**The results revealed a changing tide in software tools in fusion and fission, with more and more computational scientists preferring modern programming languages, open-source codes, and modular software. These trends represent a peek into what will happen 5 to 10 years in the future of nuclear engineering. Since the majority of our respondents belonged to US national labs and universities, these results hint at the most cutting-edge trends in the industry. The insights included in the State of Computational Science in Fission and Fusion Energy indicate a dramatic shift toward multiphysics codes, a drop-off in the use of FORTRAN in favor of more modern languages like Python and C++, and ever-rising budgets for code development, at times reaching $50M in a single organization.**

**Our survey paints a future of nuclear engineering codes that is modular in nature, small in terms of compute, and increasingly prioritized by organizations. Access to our results in web form are available [online](online).**

## I. INTRODUCTION

Computational science has long played a vital role in the design and development of fission and fusion energy systems. Computations for developing nuclear energy and weapons technologies were originally performed in cogwheel calculators, and many nuclear scientists in the 1940s did not consider electronic computers necessary to their work because computers were expensive (Kowarski, 1971).[1]

By the 1950s and 1960s, though, scientists and engineers in the nuclear field became believers of electronic computers, and early computational methods that were originally performed in a cogwheel calculator, like algebraic equations, differential equations, partial differential equations, matrix problems, and complicated functions, had already rapidly transitioned to benign performed in electronic means. By the time discrete ordinates methods matured, computers had become essential to perform the large-scale calculations needed for neutron transport problems.[1]

---

[1] In this paper, we utilize the term "fission and fusion" instead of conglomerating both terms under "nuclear." In the context of computational science, the terms 'fusion' and 'fission' are often treated separately due to the distinct scientific principles, technologies, and challenges each presents. While both fusion and fission are part of the broader nuclear energy landscape, the use of the term 'nuclear' to describe both can unintentionally conflate these two very different processes, especially in the eyes of those working in the fusion community.

The evolution towards electronic computing (and later on digital computing) was closely mirrored by the advancement of computation. As hardware has become smaller and more distilled into operating systems, humanity has been able to achieve more with computation. For example, Moore's Law, the principle observed by Gordon Moore that the number of transistors in state-of-the-art integrated circuits (units per chip) increases exponentially, doubling every 12–24 months, has been mostly true, at least until 2013. [2] The same has happened for nuclear engineering. As computation has been able to do more with less hardware (transistor density increased at least 10x every six years), fission computational science, in particular, has seen the rise in multiscale computations to achieve efficient calculations with the large number of unknowns needed to represent modern reactor cores. In particular, "using the increased power provided by multiple interconnected nodes, future reactor analysis methods shall efficiently use parallel computing tools in an integrated and consistent multiphysics environment." [1]. A good example of this phenomenon are exascale multiphysics nuclear reactor simulations, which as of 2023 have "demonstrated the capability to transport upwards of 1 billion particles per second in full core nuclear reactor simulations featuring complete temperature-dependent, continuous-energy physics on Frontier". [2] Additionally, Kord Smith has indicated in past talks at NEA that

> "[...]the challenge for future Monte Carlo simulation as the calculation of the local power of each of the fuel pins in a fuel assembly when subdivided in 100 axial and 10 radial zones for burnup calculations. The number of fuel pins in a typical fuel assembly of a PWR core is between 200 and 300 while the number of fuel assemblies in a reactor core is around 200. This results in the total of perhaps 40-60 million tallies. For an acceptable result, Smith specified that the standard deviation in each local power region should be 1 % or less. In addition, Smith considered 100 different nuclides for which the reaction rate is needed, bringing the total number of tallies to 6 billion. This huge number of tallies not only poses a problem in CPU time but also in computer memory. Smith estimated on the basis of Moore's law that it will be 2030 before such a full core Monte Carlo calculation could be done in less than one hour on a single CPU.[3]

In the context of the ever-increasing need for computational power to propel fission and fusion engineering into the future, it has become more important than ever to study how computational scientists in the industry are using and evaluating tools. Fission and Fusionprofessionals are seeing a renewed interest in energy technologies to meet the needs of data centers, with companies like Microsoft, Amazon, and Google announcing new and upcoming investments in the last year. With that new challenges for computation have arisen, including a need for advanced simulations, multiphysics software, and compounding processing needs..

Underlying all of this is the reality that the characteristics of computational tools influence what can be done with them. As [4] explain: "the materiality of technology influences, but does not determine, the possibilities for users." And so, while there is space for the computational technologies utilized in nuclear engineering to *not* determine the outcome of what is designed with them, there is probably not an insignificant amount of influence exerted on said outcomes. The tools used in nuclear energy will also become even more important as interest in fission and fusion continues to rise [5] [6] [7] and the tools used to shape such technology can shape the technology itself. This has also been referred to as the "affordances" that a tool *affords* to the wielder of said tool.

To that effect, we undertook the first ever survey of computational nuclear scientists and engineers, with the aim of uncovering their demographic characteristics, the nature of the problems their tackle, the organizations that employ them, the tools they utilize, and the degree of satisfaction with said tools. Our hope is for the results of this survey to serve as the first step in a broader, forward-looking inquiry of the nuclear energy field by uncovering the present of how computational science is being experienced and applied in the fusion and fission fields. In many ways, the way computational science is applied today to nuclear engineering will affect the energetic future of society as a whole.

The main contributions of this paper include:
- The first-ever breakdown of the demographic of computational scientists working in fusion and fission.
- A look at the main trends to come in 2025: the rise of open-source software, the rapid adoption of multiphysics software, and the coupling of AI with fusion and fission codes.
- The main factors computational scientists consider when choosing a programming language or a code.
- The average budget for development of new fission and fusion codes.

## II. BACKGROUND

The evolution of nuclear science has closely mirrored (and in part owes its pace) to the evolution of computation. Lew Kowarski, a physicist credited with outlining the design for a nuclear fission reactor in 1941, and later on founding CERN in 1953, described it succinctly: in the early 1950s, when computers were still looking for buyers,

> "...we were drawn in pretty quickly. One simple reason: the [sic] computer development took money. It started in Princeton, then it was taken up by Remington Rand (not IBM yet at that time!) but "the atom" was the wealthiest prospective user, and so the first user-motivated computers were built in places like Argonne, and in particular in Los Alamos, where Stanislas Ulam and his pupils played a considerable role in these early beginnings. Occasions for using computers gradually spread all over nuclear science. At first, they arose only in the nuclear scientist's office, at his blackboard, where theories are made and leisurely calculations are performed before and after the experiment. Not yet at the moment when the scientist is in actual contact with the nuclear phenomenon, not actually in the laboratory or at the nuclear factory. At this stage, the computers are useful for what might militarily be called staff work, not for field work." (Kowarski, 1971. p.208)

Like a hammer looking for a nail, computers seemed poised to solve problems for nuclear scientists that they perhaps didn't even envision as such. The best example is the fact that, initially, scientists did not even consider computers to perform their fission research duties. As the utility of computers became clearer, they began to be used in the research laboratories themselves, then on to the manufacturing machines utilized to create reactors and factories that produced said machines, and finally even to the libraries where the engineers and scientists were learning how to do their work. By 1971, Kowarski described the machines as being everywhere, with "no place to hide." (Kowarski, 1971. p.209)

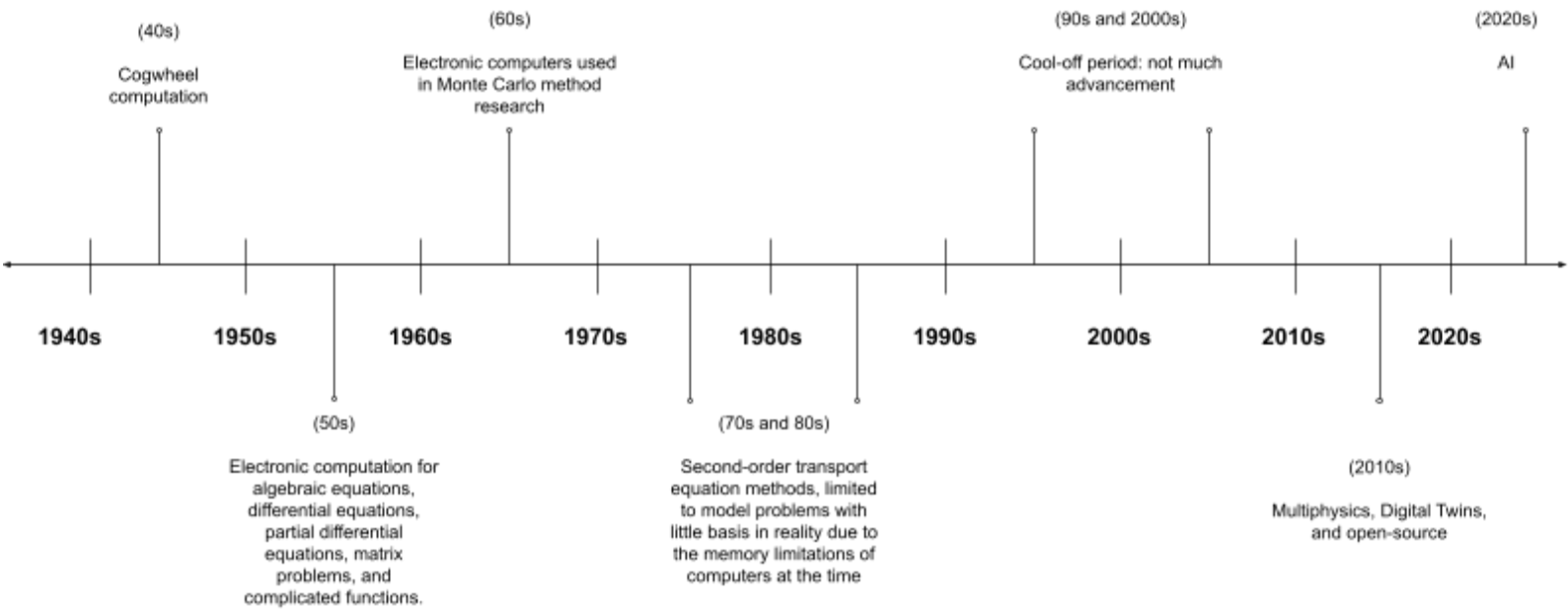

The co-evolution of general computation and fusion and fission computation

With their rapid introduction, computers became ever more entrenched in the work of the nuclear engineer. Pretty quickly, engineers and physicists in the field understood that they could mathematically do with computers what would otherwise be incredibly slow, difficult, or downright impossible. For example, before computers, to estimate the integral of a function over a region, scientists would pick random points in the region manually (perhaps using dice or some other method of generating random numbers). One would then calculate the function's value at each of those points. Arriving at a good estimate, required repeating the process tens of thousands of times, manually performing calculations or using simple tools like calculators. This process could take days or weeks, and was limited by the ability to track large numbers of random samples. Withcomputers, random samples could be generated within milliseconds, and the computer could quickly compute the function and results. In other words, the same task that might have taken weeks before can now be done in seconds or minutes on a modern computer by the 40s.

By the 50s and 60s, John von Neumann and Nicholas Metropolis, who were once involved in the development of the first atomic bomb, and later on involved in the development of Monte Carlo methods, utilized computers to continue researching Monte Carlo, making the most of the speedier and more memory-rich mainframes of the time to accelerate nuclear calculations. [1]

By the 70s and 80s, second-order transport equation methods began to flourish due to the advances in computational power, though they were limited to model problems with little basis in reality due to the memory limitations of computers at the time. Since then, as [1] (2010) also mention, "It has only been with the rapid expansion of computer memories since that time that second-order methods have become a viable alternative for engineering calculation, and production codes based on the methods became viable for large-scale engineering computations." (p.86) This reflects the continuous trend of co-evolution of computational science and core nuclear science, and the deeper correlation between tools and what can be done with them: sketches of theories and methods arose, only to be proven too hard to develop, but as computation became more powerful, they were utilized to implement said theory, catapulting science forward, as they did with Monte Carlo methods.

The 1990s represent a cooling stage for nuclear energy as a whole in the USA which had been largely at the forefront of computational advances. Though computation was developing by leaps and bounds, Hwang, in [1] describes that time period, from the 90s all the way until the first decade of the 2000s, as one of uncertainty and relative silence in nuclear computational science:

> "There is no question that nuclear developments in the USA have suffered. It is unfortunate that the lack of funding also reduces the traditional interest within the nuclear community in the pursuance [sic] of rigor and better understanding of the basic concepts essential to the future of nuclear energy. The dilemma is that a great many difficult issues encountered previously could probably be resolved today given our better knowledge of nuclear data, improved methodologies, and the availability of efficient but less costly computational facilities." (p. 210)

This brings us to the two most recent decades. The pace of computational innovation in nuclear has picked back up, with eyes increasingly turning to the higher-fidelity possibilities of simulations and integrated systems made possible by the ever-increasing computational power hosted in even smaller computers. A renewed interest in nuclear energy, along with the mentioned advances in computation, has birthed new exciting concepts: small modular reactors, multiphysics software, autonomous reactor operations, Digital Twins, Artificial Intelligence(AI)-aided simulations, and synthetic data generation for said simulations.

For example, the development of the Multiphysics Object-Oriented Simulational Environment (MOOSE) at the Idaho National Lab started in 2005 as part of an initiative to build a multiphysics computational framework for nuclear power applications.[8] But it wasn't until 2012 when MOOSE began to be utilized publicly by computational scientists in the field, outside of the Lab itself. Almost concurrently, OpenMC, a Monte Carlo software focused on multiphysics coupling in an open-source setting, began to be developed in 2011 by scientists at MIT [9][10]. By 2018, MOOSE had 1000 users [11], and OpenMC currently has hundreds of contributors to its open-source repository. [10] Additionally, the rapid pace of adoption seems to be mirrored by the rise in job posting for OpenMC knowledgeable computational scientists. In total, since 2022, 47 jobs have been spotted and posted on the forum.[12]

With the advent of the second decade of the 21st century, computational nuclear scientists and engineers are now turning to AI. In particular, computational scientists are galvanized by the potential for AI to aid with the design of fission and fusion energy systems, fuel management, thermal-hydraulic simulations, and radiation shielding design, among other common problems. [13] However, questions remain about the lack of experimental data and the increased risk of data distribution drift and imbalance, as well as the black-box dilemma where AI models cannot explain their methods, leading to hallucinations and poor interpretability. [14]

In the context of these new horizons in computation for fission and fusion, the authors of this paper consider it of the utmost importance to more closely investigate the nature of computational fission and fusion science and scientists. Our interest in the area is rooted in a desire to understand how the tools that designers use shape their design choices and processes.

## III. TERM DEFINITIONS

Several terms are used throughout this paper that need proper definitions to avoid confusion. Here's a table detailing them:

| *Term* | *Definition* |
| --- | --- |
| Affordance | The degree to which a tool influences the outcome of what is designed with it. |
| Computational Science | Computational Science is the study and application of computational methods and algorithms to solve complex scientific problems that are difficult or impossible to solve through traditional experimental or analytical approaches. It combines elements of computer science, mathematics, and scientific computing to model, simulate, and analyze phenomena across various disciplines like physics, biology, chemistry, engineering, and social sciences. |
| Developer Experience (aka DevEx) | Developer Experience refers to the overall experience and satisfaction a developer has while interacting with tools, technologies, and platforms |

|  | they use to build software. It encompasses everything from the ease of setting up and using APIs, libraries, and frameworks to the quality of documentation, support, and the design of development environments. |
|---|---|
| User Experience (aka UX) | User Experience refers to the overall experience a person has when interacting with a product, system, or service, particularly in terms of how easy, efficient, and enjoyable it is to use. UX encompasses all aspects of the user's interaction, including the design, usability, functionality, accessibility, and performance of a product. It involves understanding the needs, behaviors, and emotions of users, and designing experiences that meet those needs in a seamless and intuitive way. |
| User Interface (aka UI) | The visual elements and interactive components of a product or system that allow users to interact with it. This includes things like buttons, icons, menus, sliders, text fields, and any other elements that users engage with to control or navigate the system. |
| Code | A nuclear code (in the context of fusion and fission) typically refers to a set of computer programs or algorithms used for simulating or analyzing nuclear systems and processes. These codes are used for a wide range of applications in nuclear engineering, including reactor design, nuclear safety analysis, radiation transport, fuel cycle modeling, and fission/fusion simulations. |
| Programming language | A programming language is a formal set of instructions used to communicate with a computer and control its behavior. It provides a way for developers to write software, algorithms, and applications by specifying operations, data structures, and control flow in a syntax that both humans and machines can understand. Examples include Python, Java, Rust, C++, and many others. |

A further extended glossary of useful terms is included in Appendix 1.

## IV. METHODS

To develop a comprehensive view of the current state of computational science in fission and fusion, we created a survey that was distributed to researchers and professionals in the nuclear field.. The survey had 25 questions, divided into the following sections:
1. Personal characteristics:
    a. Basic categorization: determining if a respondent works in fusion or fission, and their area of specialization
2. Problems being solved with codes, at each respondent's workplace: categories of problems being solved with software
3. Organizational characteristics: size of organization they work in, category of organization, years of experience in said organization.
4. Software tools: programming languages preferred, codes utilized most often.
5. Single code feedback: respondents were given the option to choose one code to focus on to give feedback on what it does correctly and what it could improve.
6. Code wishlist: factors that make for especially good codes in the eyes of the respondents.

The questionnaire was developed after an initial set of 3 informal interviews with industry professionals, researchers, and professors.

Readers can see the full set of questions of the survey in the appendix of this paper. The questions were developed iteratively, with the initial questions being informed by informal interviews with nuclear software developers and users, followed by input from industry professionals and computational nuclear science faculty. The survey design was inspired, partly, by the renowned StackOverflow State of Development Annual Survey [15], which asks questions in a similar set of categories to the ones we have outlined above, but in the broader context of field-agnostic software development.

The survey was disseminated through social media posts on the Fastest Path to Zero FB and Twitter accounts, a publication on University of Michigan's Nuclear Engineering & Radiological Sciences Departmental Newsletter, emails from original reviewers of the survey to their contacts in computational science, and emails from University of Michigan professors to their professional and academic circles.

In total, our survey remained open from June 27, 2024 to September 1, 2024. By the end, 103 computational scientists responded.

We note that certain questions were not answered by all 103 respondents. Though responses were mandatory for most questions, in cases when the data seems to be insufficient, there are two potential causes: certain respondents did not finish the survey, or there were issues with browser cookies or partial completion settings on the respondent's part.

It is logical to assume that, at present, there might be no more than a few thousand computational scientists working on the development of fusion and fission codes, all across the world. Regardless, there are no scientifically accurate accounts of the number of computational scientists working in fusion and fission, so we aimed for a statistically non-representative stratified sample[16], aiming to reach a broad range of software developers in the fission and fusion field working across different codes and programming languages.

We consider this survey to be a first glance at an area of work that is vital to nuclear science and engineering but whose landscape is not well understood. . The survey we carried out as part of this study constitutes the first round or benchmark and we aim to repeat the survey regularly – likely every year.

The rest of this paper is organized in sections V-XII. Section V presents the results, subdivided into responder profile, problems being solved by respondents with computational science, the tooling and assigned resources respondents have available to them, and the developer experience of such tools. Section VI is the discussion of the survey's results, Section VII expounds on the study's limitations, Section VIII presents conclusions, IX suggests future work in the area, X the references, XI acknowledgments, and XII appendices.

# V. RESULTS

The results section is divided into several subsections. The first one pertains to the overall responder profile: the type of areas of expertise of the respondents, and the characteristics of the organizations they work in. The second subsection goes further into the types of problems that are being solved by respondents with codes. The third subsection deals with tooling and assigned resources: the budgets available to the computational scientists, the codes they use most and least, and their programming languages patterns of usage. Finally, the last subsection asks respondents about their experience as developers: what makes for a good code, or a bad code, and their opinions on why.

## V.I Responder Profile

Figures 1 and 2 below show how the respondents are distributed across fission and fusion and sub-fields. 54% of respondents reported working on fission applications online, 23% on fusion only and 24% on both fission and fusion.

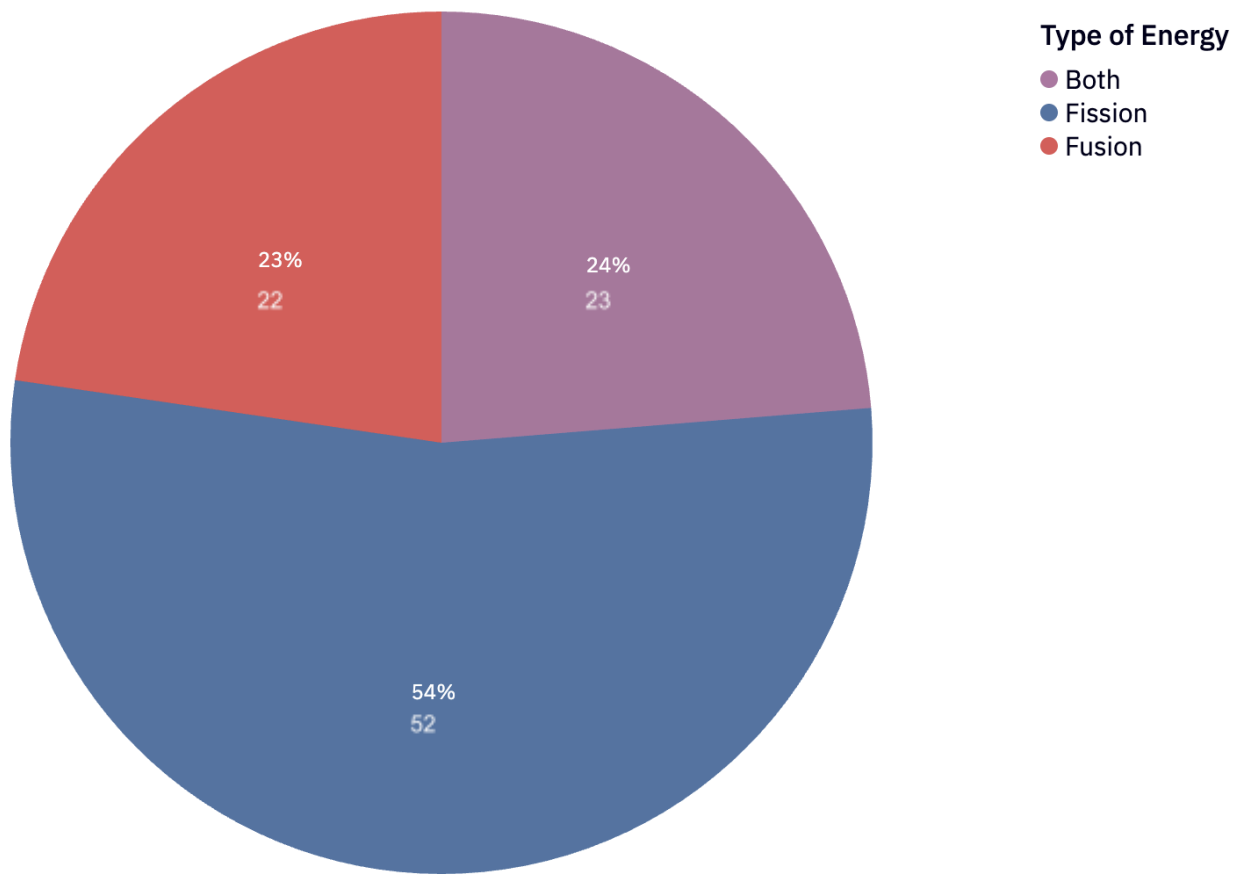

*Fig. 1: The majority of survey respondents work in fission.*

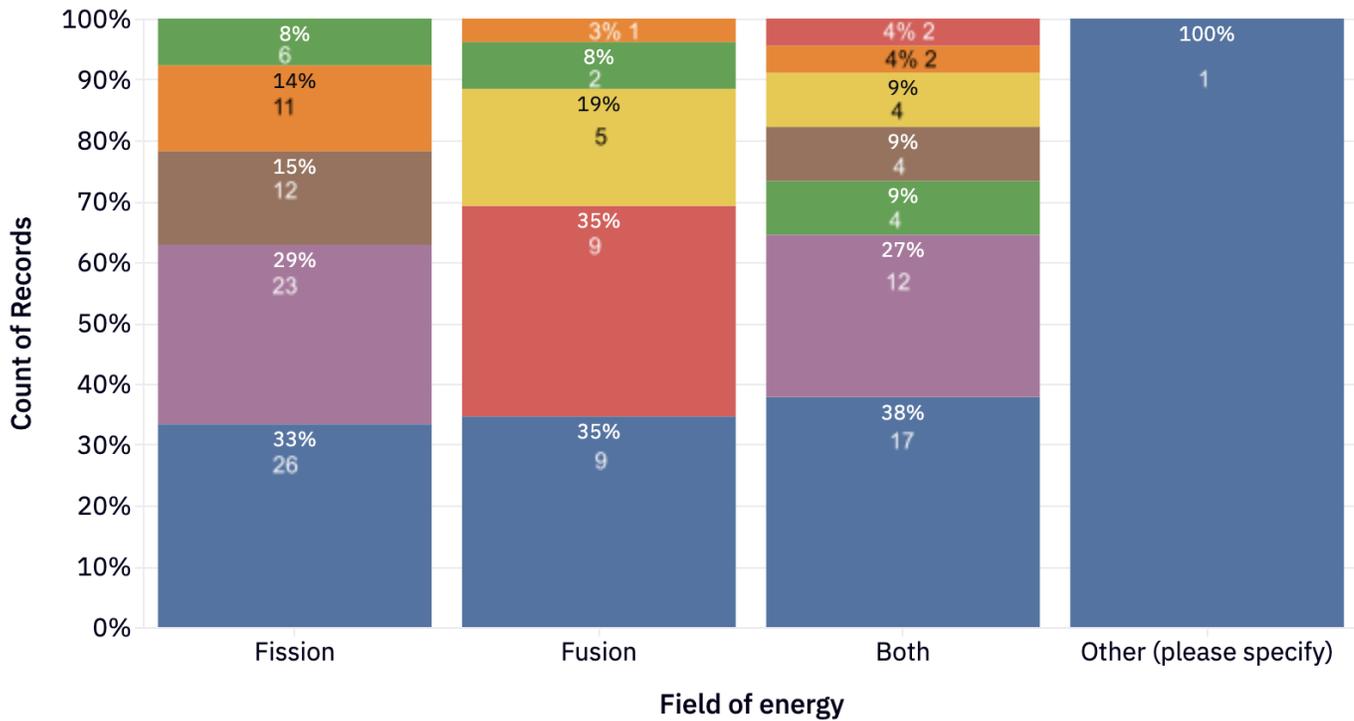

*Fig. 2: The area of specialization for most respondents was computational science.*

Figure 2 shows a further breakdown of the results in Figure 1 by sub-fields and areas of application within fission and fusion. The majority of all respondents work in computational science, specifically. In the case of fusion, participants are split 35% working in computational science, and 35% in plasma physics. In the case of fission, 33% of participants were specialized in computational science, followed by 29% in reactor design.

In the Other category, the remaining responses specified reactor physics, fuel performance, neutronics analysis, neutron transport, health physics, uncertainty quantification, HTS magnets, and transport theory and methods. Computational weapons/stockpile stewardship-related work was deliberately not included due to the sensitive or classified nature of that work, so we expected few (if any) responses from that group.

Figure 3 shows the breakdown of respondents based on the size of the organization where they work. The vast majority of survey respondents (78%) reported having worked in their field for over 5 years, and the organizations they work at tend to have over 500 employees (72% of all respondents). There is a strong cohort of smaller companies (17%) with 11-50 employees investing in codes, which could indicate a rise in startups interested in creating their own codes.

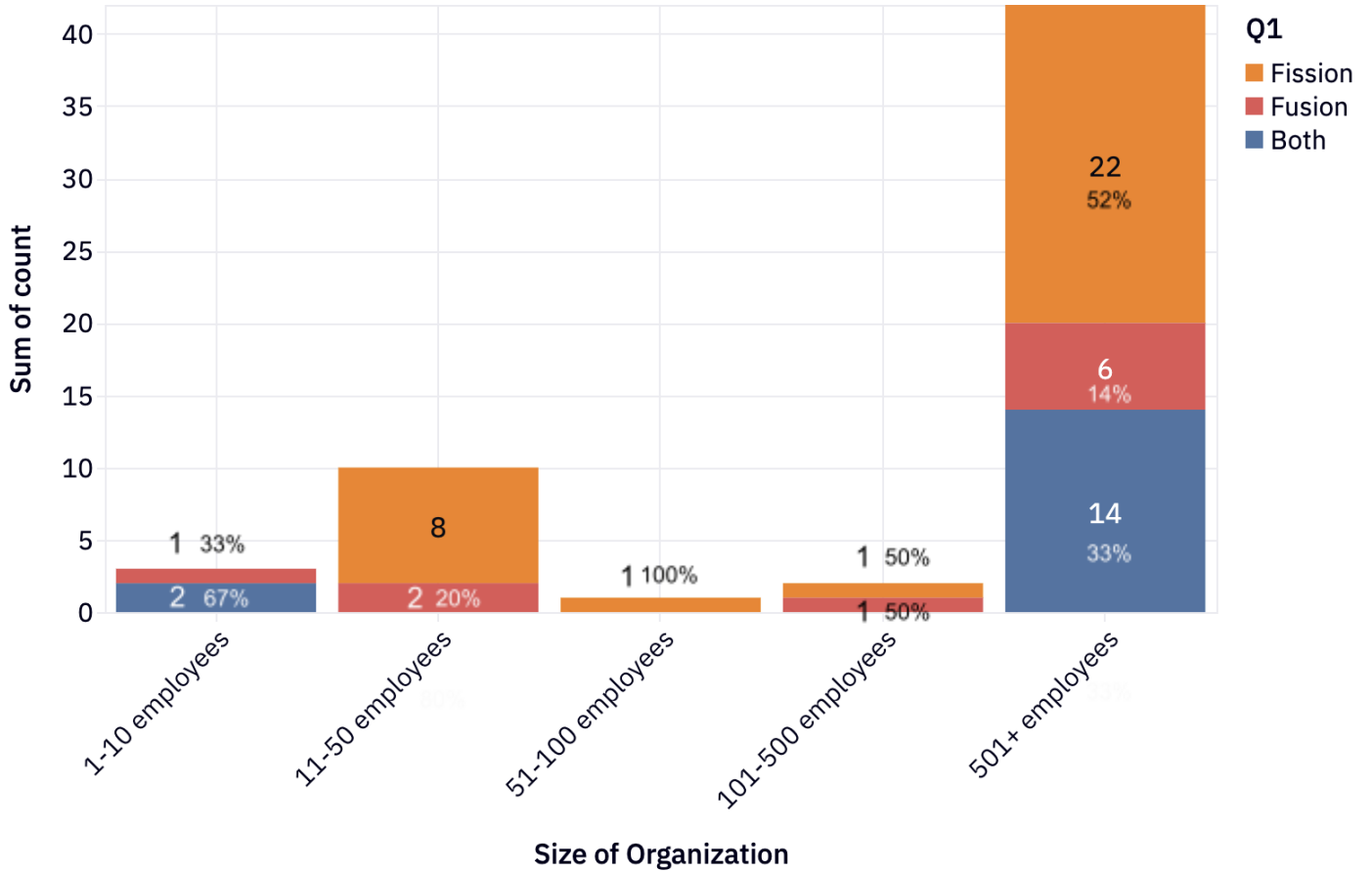

*Fig. 3: "What is the size of the organization you are a part of?"*

Figure 4 shows the distribution of respondents across different types of organizations – namely national labs, universities, and companies. It is noteworthy that none of the respondents work at electric utilities. Our survey skewed towards national labs, universities, and startups, which might explain these results. Also notable is that 70% of the respondents are full time researchers and practitioners and 30% are students.

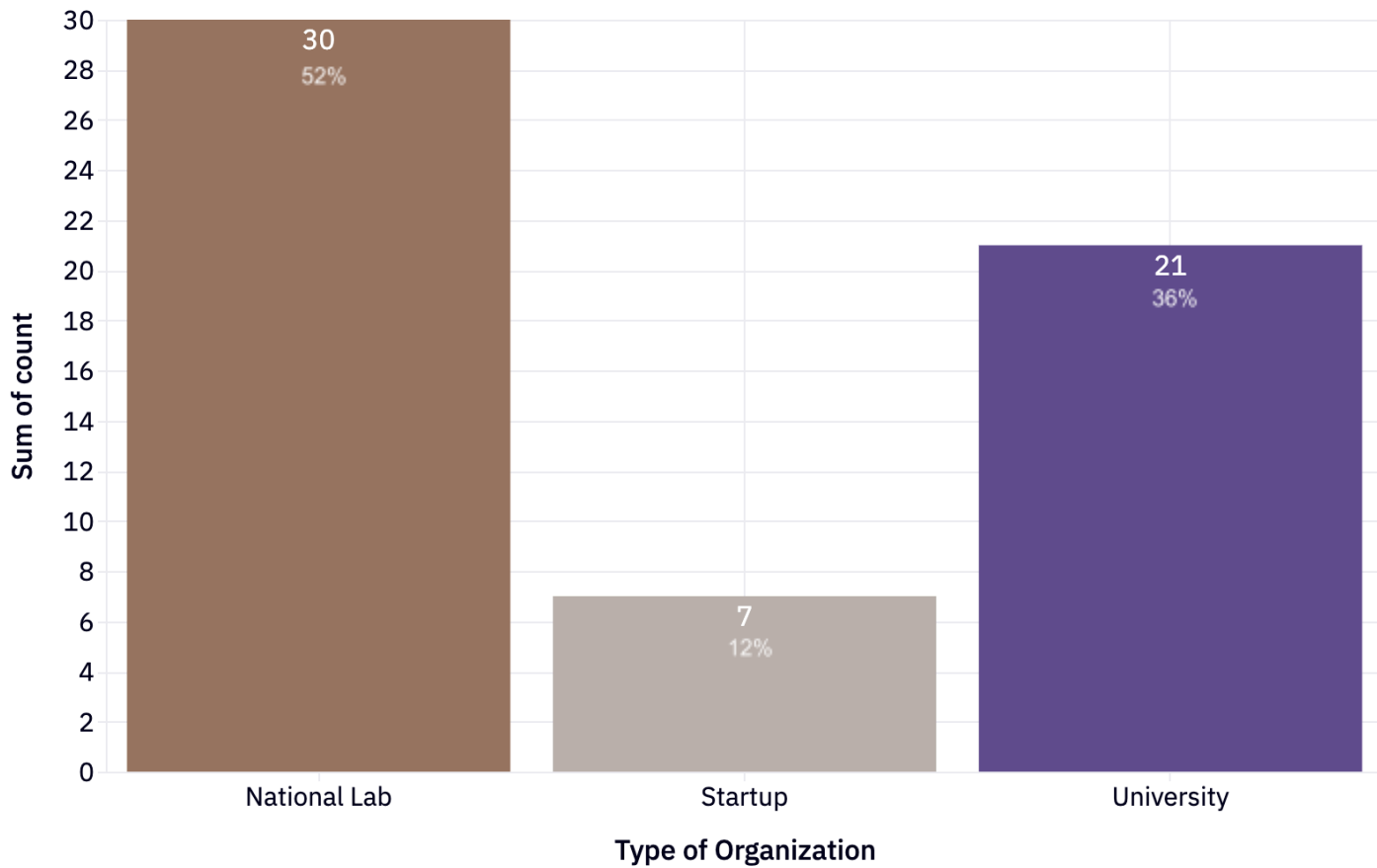

*Fig. 4: The majority of respondents were a part of national laboratories. Not all participants responded to this question due to a glitch in the system allowing them to skip it.*

Figure 5 shows a distribution of the respondents based on years of expertise in computational nuclear science and engineering. X% of the respondents reported having worked in computational nuclear science and engineering for over 10 years, [add more on the other categories]

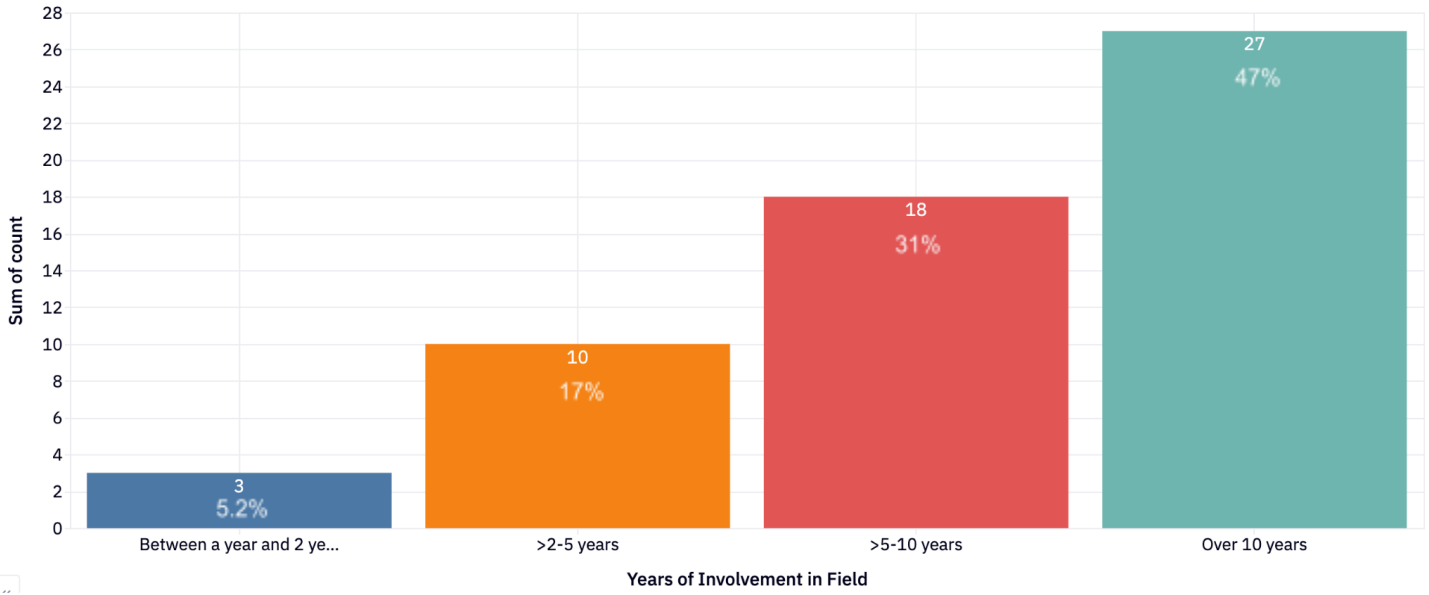

*Fig. 5: 27% of participants had been working in computational science for over 10 years.*

Figure 6 shows that more than 82% of all respondents use codes as a daily part of their jobs, indicating that computational scientists tend to not only program codes (potentially), but also use them actively. More research is needed on whether the makers of codes themselves also use them daily, or if there is a divide between writing codes, and using them at work.

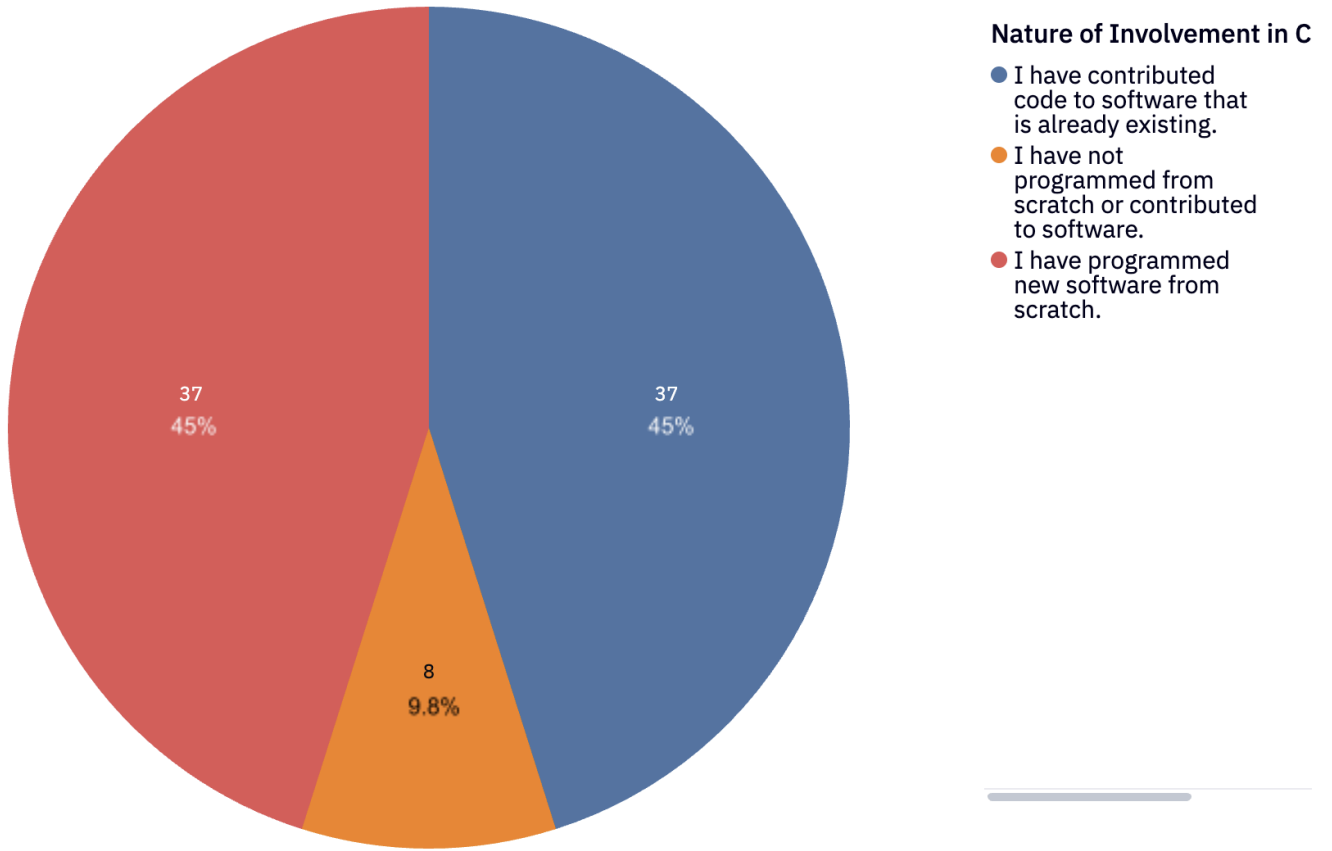

*Fig. 6: The two major natures of involvement in code creation are the contribution to new codes and the contribution to existing codes.*

Figure 7 shows the intended uses of codes created by the respondents. We asked this question because we were interested in learning the intent behind creating nuclear software – is the software created directly for use by the person creating it, is it created for use within the respondent's organization, or is it created for use outside the home organization?

Interestingly, in the simple majority of organizations where codes are created, 10 people or more tend to use them. The nature of those users (is it the team of computational scientists?; is it some other type of audience?) remains to be explored in future research.

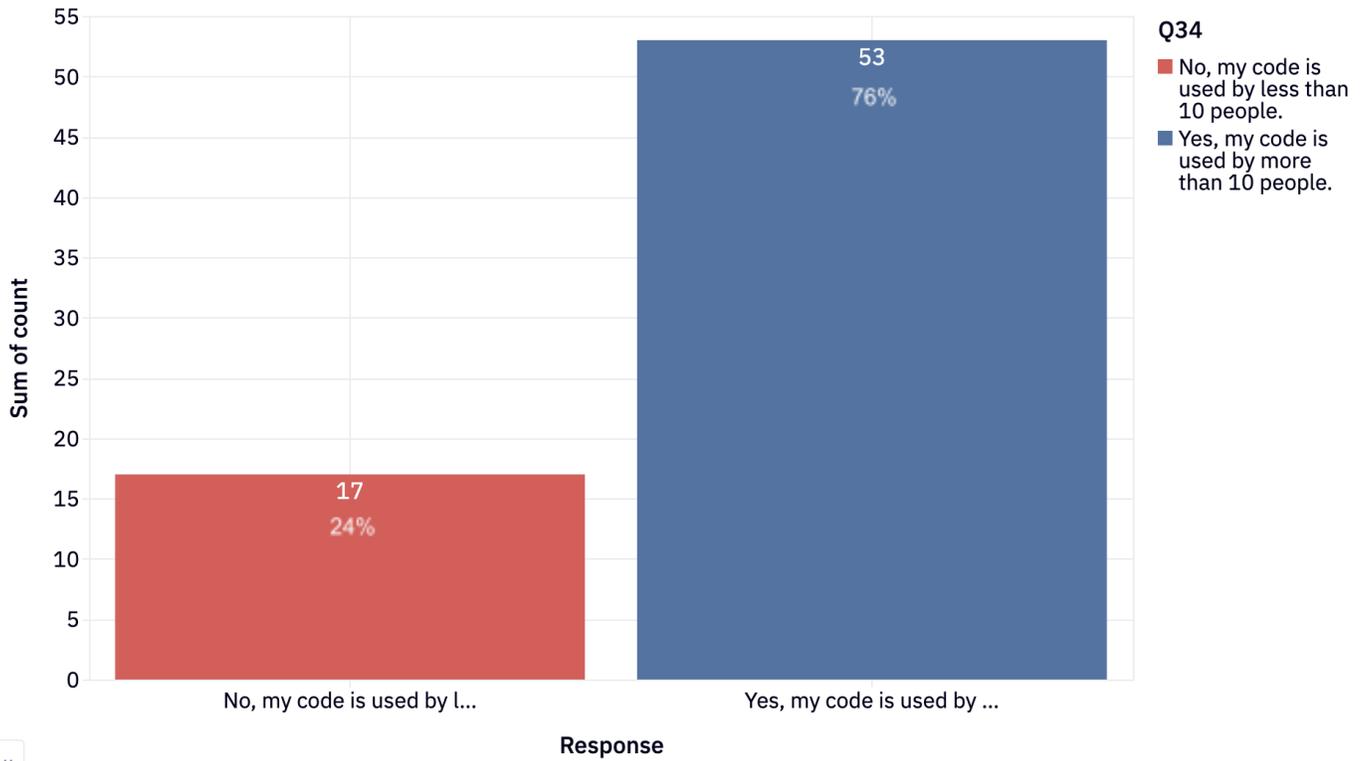

*Fig. 7: The vast majority of respondents that created codes or contributed to them have at least 10 people in their organization use them*

**V.II Problems Being Solved**

In this subsection we explore what problems are solved using nuclear software. 25% of respondents are utilizing codes to perform parametrics studies for design, 24% focus on safety analysis, 8.5% parametrics study for numerical methods, and 1.7% on uncertainty analysis. Interestingly, a large number of respondents (41%) chose the "other" option, indicating their uses were not captured in the options provided in the survey.

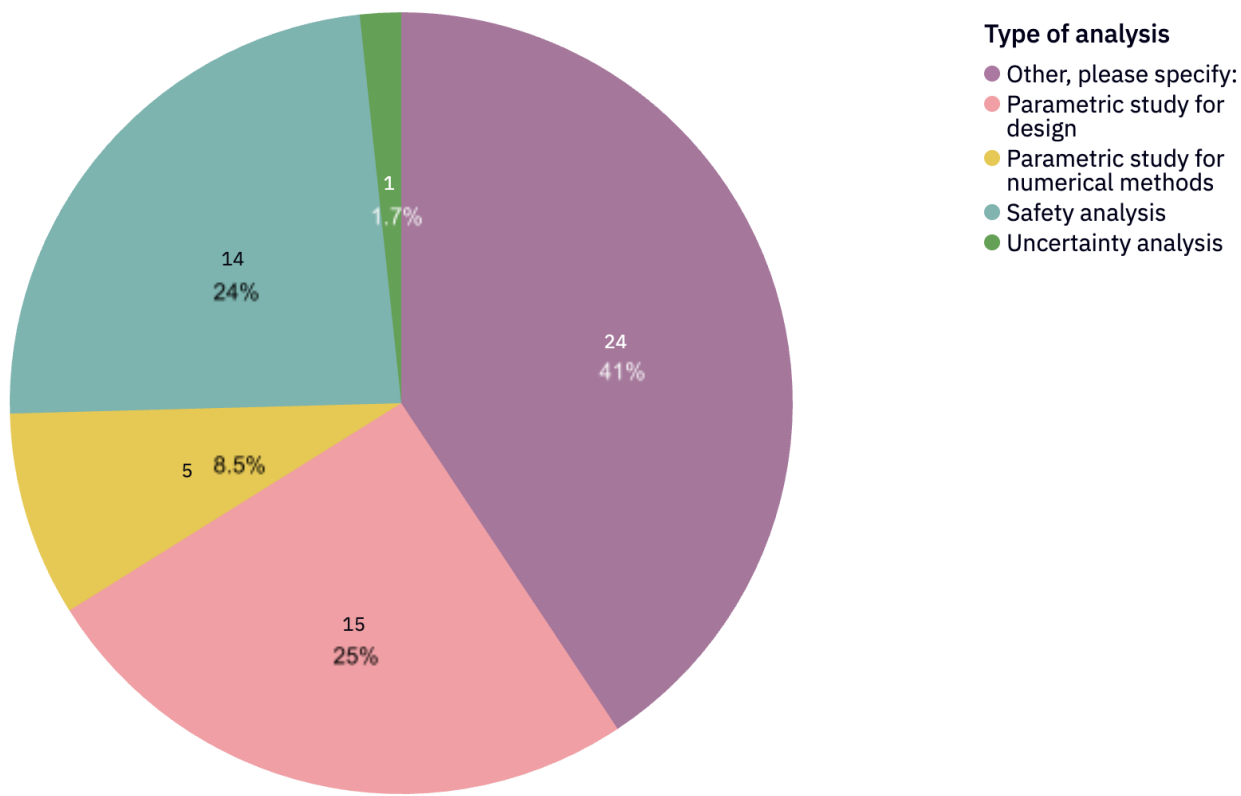

*Fig. 8: 59 respondents chose a specific type of analysis performed with codes.*

When respondents self-reported "Other, please specify" as their type of analysis, 7 mentioned working on "all of the categories", and 2 more were specifically focusing on multiphysics and multiscale modeling. Other responses were empty or indecipherable.

This trend continues when looking at the types of physics computational scientists are focused on, since multiphysics is favored by a majority.

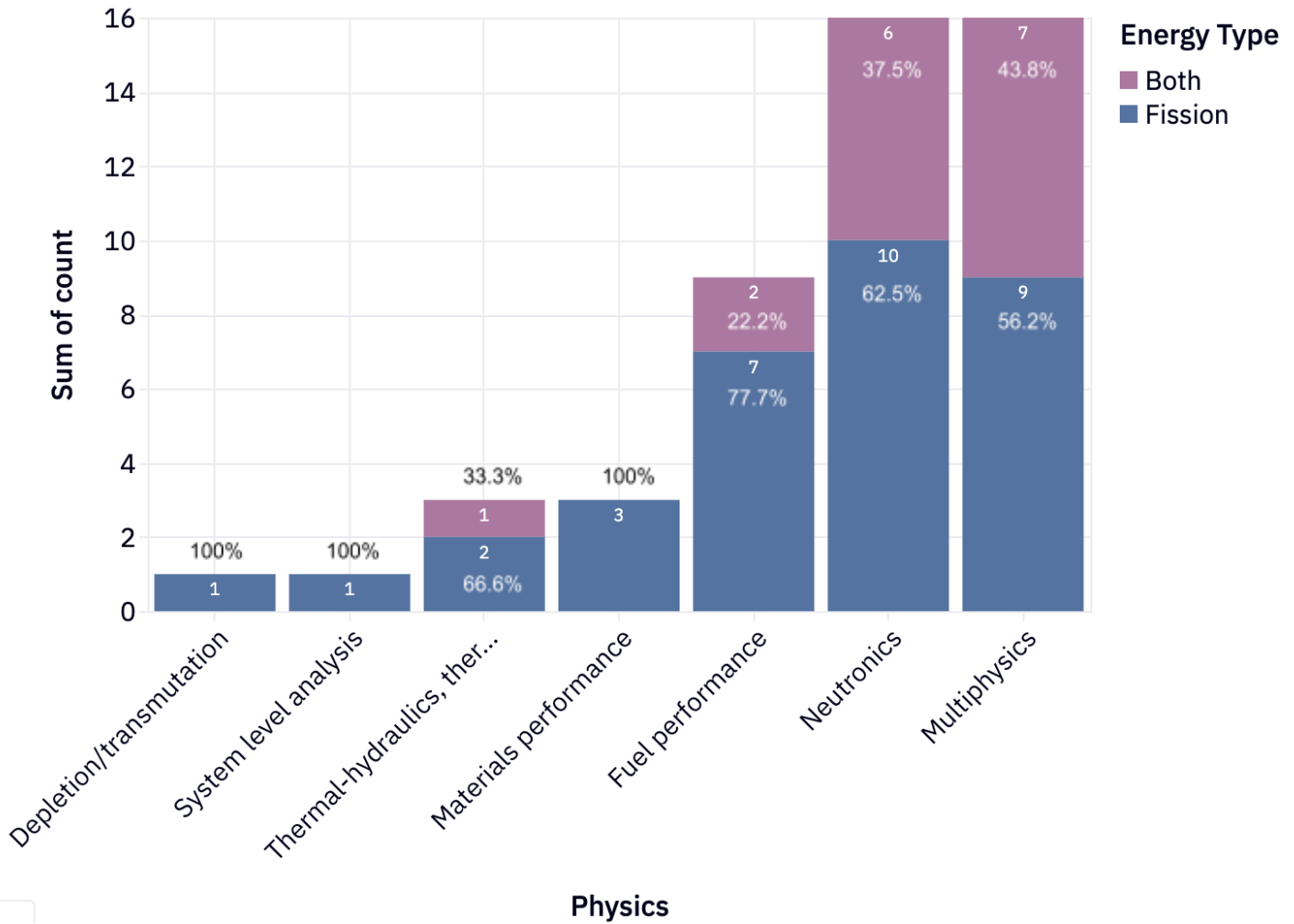

Fig. 9: Multiphysics and neutronics are the focus of most codes developed by computational scientists

In general, the problems being solved by computational scientists can be summarized in Table 1 below. Quotations come directly from respondent answers.

| 1. Multiphysics and Multiscale Modeling | 2. Reactor and Device Design and Optimization | 3. Neutronics and Radiation Transport |
|---|---|---|
| Many respondents mentioned the study of multiphysics interactions (e.g., radiation transport, thermal-fluids, plasma physics) in fusion devices. Examples include:<br><br>● "Multiphysics interactions between radiation transport, | Several responses emphasized the development of fusion device design, including power plant design, plasma control systems, and divertor design. Examples:<br><br>● "Fusion power plant design"<br>● "Plasma control system, data | A notable portion of the responses involved neutronics and radiation transport in fusion systems, especially in terms of shielding, neutronics calculations, and neutron transport simulations. Examples: |
| thermal-fluids, and fuel performance"<br>● "Multiscale modeling of fission and fusion reactors"<br>● "Understanding multiphysics and multiscale mechanisms at the intersection of plasma, neutronics, materials, and engineering processes"<br><br>There is also a focus on multiphysics modeling in fission reactors, often involving thermal-hydraulics, neutronics, and fuel performance. Examples:<br>● "Solution of neutral particle transport"<br>● "Multiphysics model of advanced reactors"<br>● "Dynamic multiphysics simulation framework for nuclear system analysis" | acquisition"<br>● "Divertor design, exploring fundamental physics"<br><br>For fission reactors, the focus is on advanced reactor design, fuel performance, and optimization of reactor systems. Examples:<br><br>● "Reactor design, AI/ML, control, optimization, safety analysis"<br>● "Rapid deployment of advanced reactors"<br>● "Predict the behavior of nuclear reactors under nominal and accidental conditions" | ● "Neutron transport simulations for experiment analysis and shielding"<br>● "Simulation of neutrons in fusion reactors and finding their impacts on the design"<br>● "Stellarator neutronics design and analysis"<br><br>Similarly, neutronics plays a central role in fission reactor analysis, especially for shielding performance, neutron transport, and burnup extension. Examples:<br><br>● "Efficient and accurate fission reactor analysis"<br>● "Burnup extension in LWRs"<br>● "Neutron shielding performance, shutdown dose rate calculations" |

*Table 1: Respondents are working on solving multiphysics interactions for fusion devices, as well as in fission reactors. Computational science is also being actively utilized to design devices and reactors, as well as solving neutronics and radiation transport problems.*

### V.III Tooling and Assigned Resources

First up: programming languages. In general, Python is the preferred language, with C++ coming in at a close second.

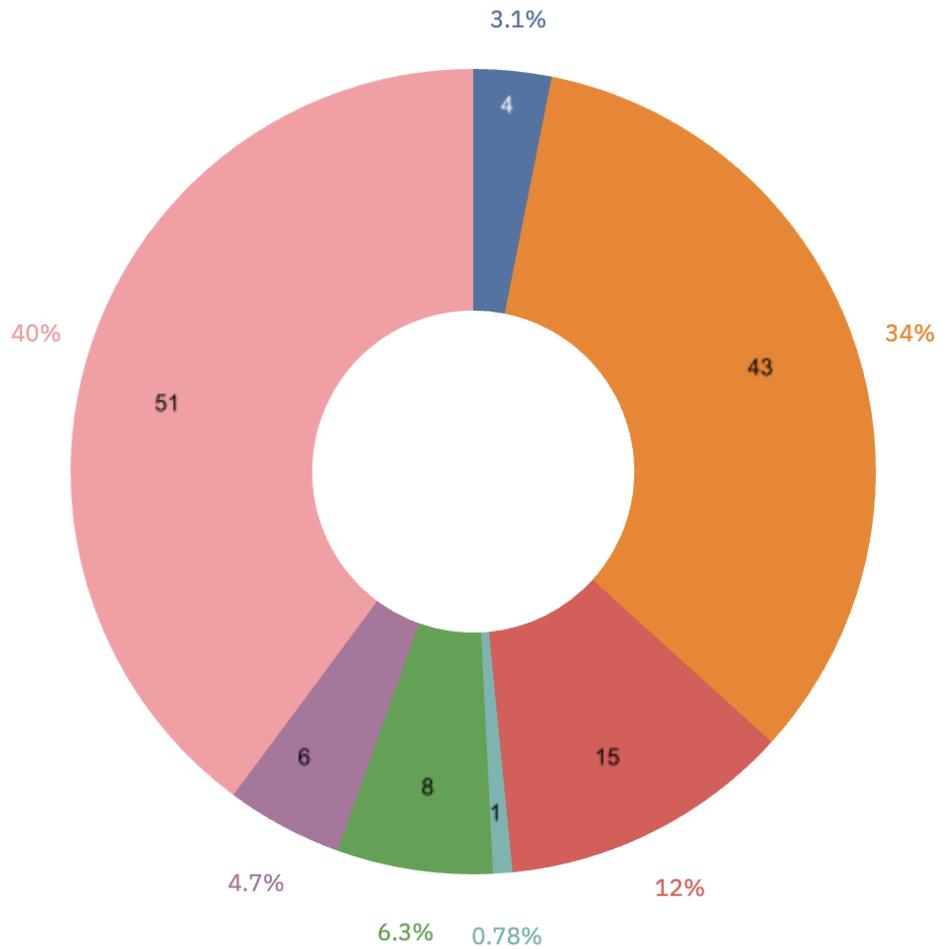

*Fig. 10: The top 3 programming languages utilized by computational scientists are C++, Python, and Fortran.*

Other languages mentioned include Mathematica (2 responses), Rust, SCALE/MCNP (in spite of not being a language, it was noted by respondents in the fill-your-own option), Javascript, and Ansible (same case as SCALE). Related to the mention of Rust, we expect a trend in the next couple of years towards lower-level languages per the White House's February 2024 recommendation. [17]

Still, Python and C++ are the leaders, and will probably continue to be so in the foreseeable future (see the desirability of programming languages in Figure 13, which described the top languages computational scientists wish to work with).

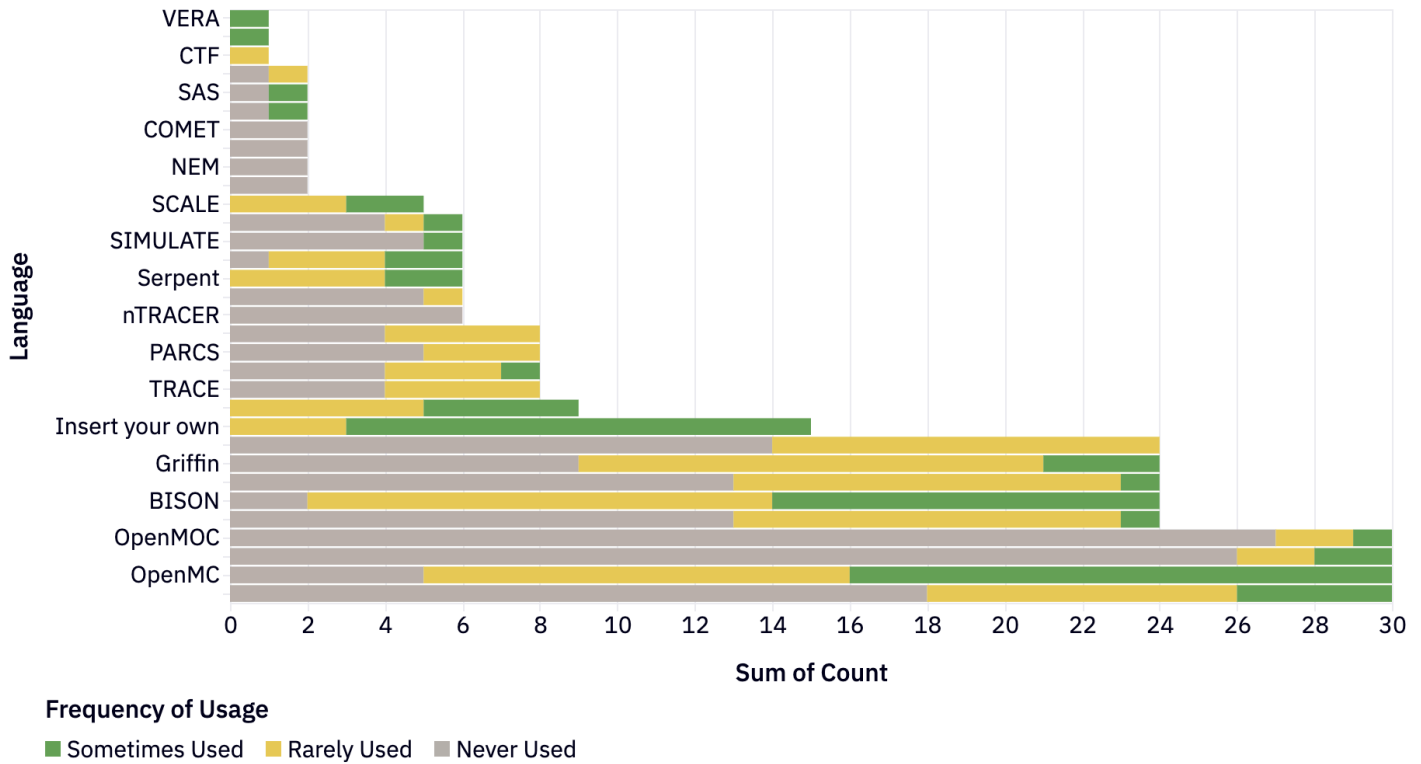

Fig. 11: The least utilized code is OpenMOC, the most utilized one is OpenMC.

Though the answers to the programming languages used by respondents might be biased by the nature of our respondent profile (a majority of which works in universities and labs that develop multiphysics codes), it matches an overall shift towards multiphysics code adoption, as reflected in the rapid rise in OpenMC forum job postings.

**Investments in code**

In more resource-related results, investment in code development can range from $0 to millions of dollars, but a significant portion of computational scientists do not have insight into the budgets managed by their teams. The median response for the estimated budget spent on nuclear engineering software is $1 million dollars. The distribution is heavily skewed, with a few large values pulling the mean upward, while the majority of responses are clustered around lower values, specifically zero.

Other important results around budget for code development are:
- The minimum budget is $0
- The 25th Percentile: $1000
- The 75th Percentile: $5,000,000
- And the maximum budget in responses was $50,000,000

As for programming languages, for what's left of 2024, computational scientists don't have curiosity for new languages, nor a particular disdain for the languages they currently use, though C has dropped off in desirability a bit, and so has MATLAB and Python.

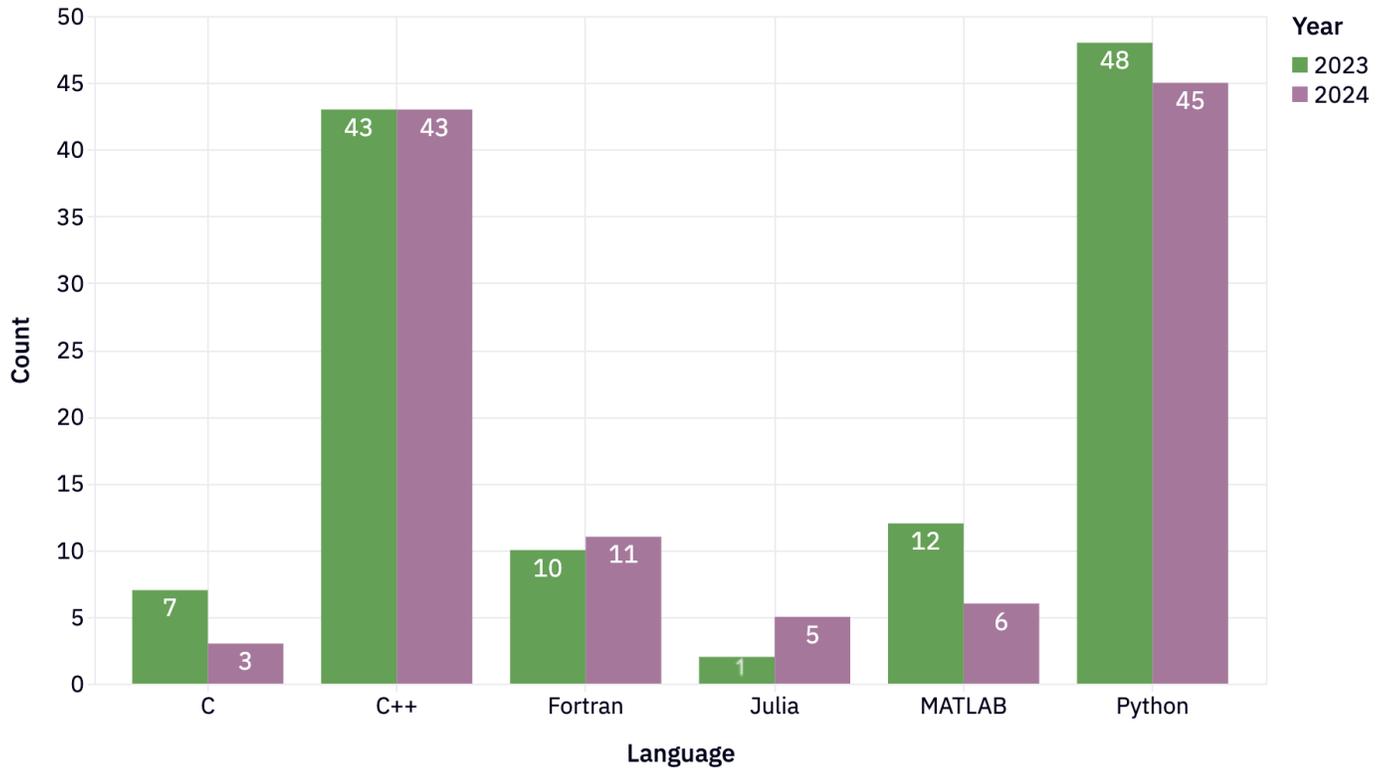

Fig. 12: Each respondent specified which programming language they used in 2023 vs. which one they wanted to work with in 2024. The difference implies potential satisfaction with the programming language used or a desire to change usage patterns.

We asked computational scientists to look at the top 6 languages in fusion and fission according to experts consulted while designing this survey. There doesn't seem to be an appetite to try out new languages this year. Though there is an increase in interest for Julia and a slight one for Fortran, the numbers were so small to make them statistically ignorable.

Finally, the codes that seem to generate the biggest split in opinion are Griffin, MCNP, and OpenMC. We did not inquire further into why, but it would be a helpful future question in semi-structured interviews, which this research team hopes to do with select respondents this year.

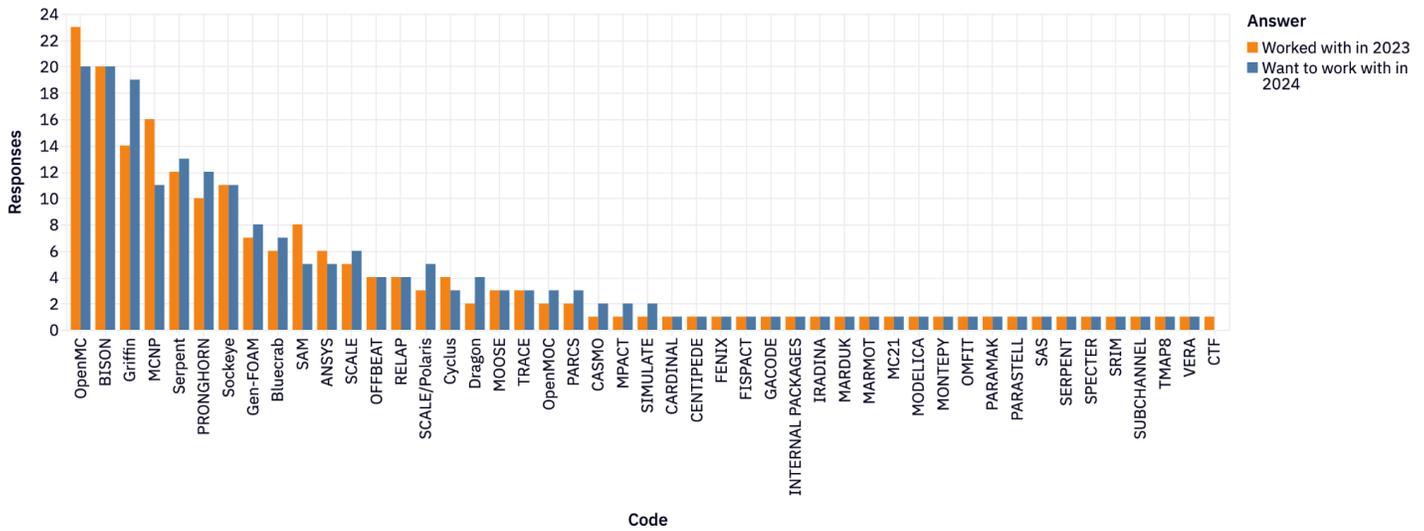

*Fig. 13: A comparison of the codes used 2023, vs the ones that scientists desired to use in 2024. OpenMC sees a rise in popularity, but a decline in desirability for 2024.*

**V.IV Developer Experience**

We also asked respondents about the usability of the codes they utilize on a daily basis. For the next section, respondents were encouraged to choose one (1) code to focus on. Before presenting the results, it's important to mention that the vast majority of the respondents did not create the code itself, which gives the results for this section of the survey even more relevance. In User Experience and usability, objectivity is preferred. Often, when a creator of a tool evaluates its usability, it is tainted by their previous knowledge of where buttons are, why they were placed where they are, what the steps towards task completion must be done, etc. It is motivating to be able to get less biased results from our respondents.

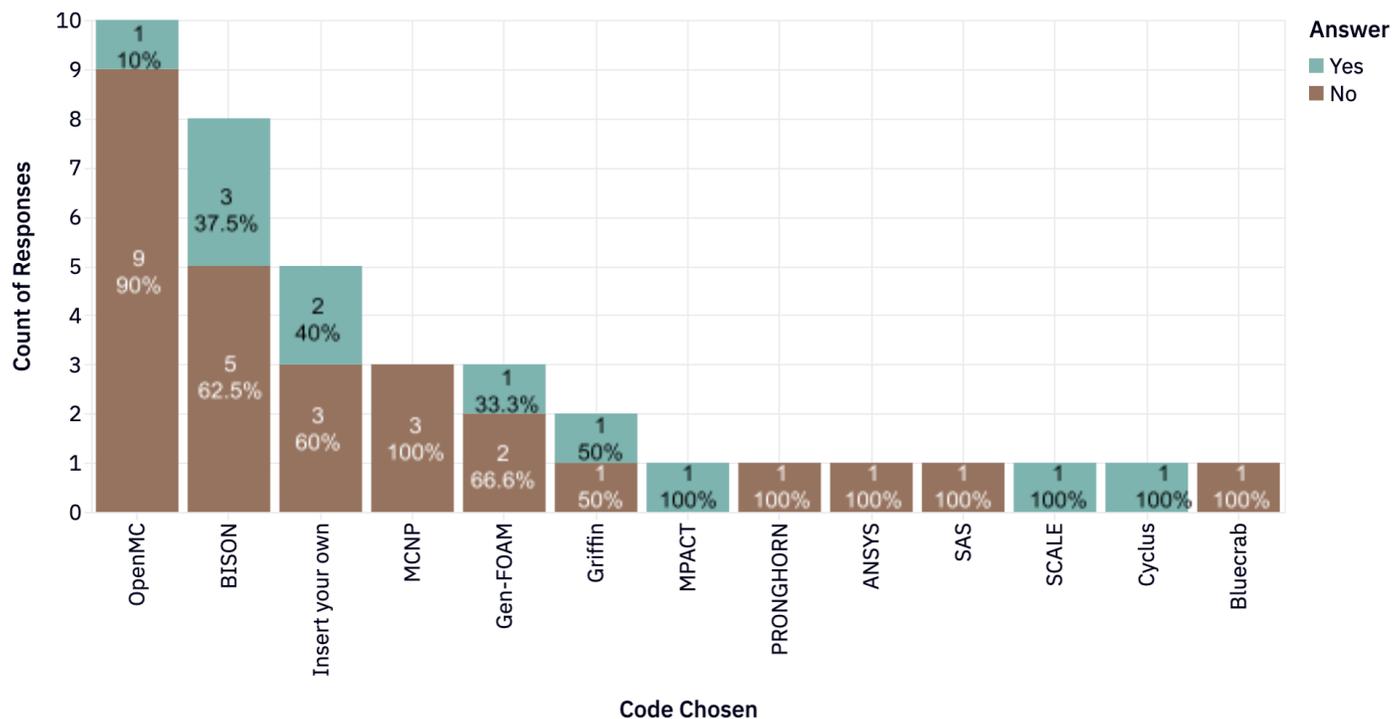

*Fig. 14: After respondents chose a code they wanted to give more feedback on, we asked them if they were involved in its creation. The majority did not, except in the case of BISON, where the results are a bit more biased.*

In a follow-up question, we asked computational scientists who were involved in deciding which code they would use for their daily tasks. OpenMC was chosen by scientists themselves, without top-down organizational pressures. BISON, on the other hand, tends to be imposed on computational scientists by the heads of their organization. This question is important to assess if the future desirability of any one code is dependent on organizational pressures, or if it is a truly emergent trend.

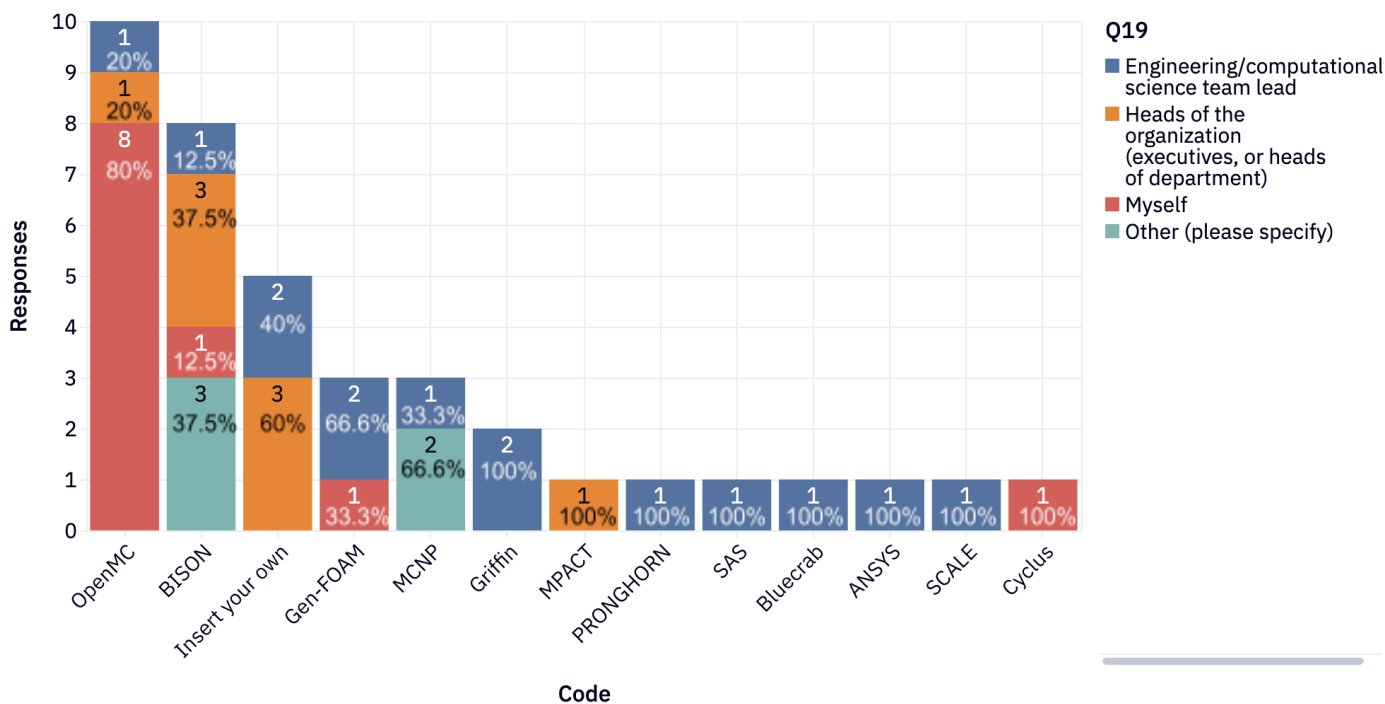

Fig. 15: OpenMC is most freely chosen by scientists for daily use, free of organizational pressures. BISON is most chosen by heads of organizations and executives.

When asked what their top 3 reasons for being dissatisfied with their code of choice are (Appendix 3), respondents suggest the lack of open-source access, slow performance, limited development activity, and unresolved errors are significant concerns for users.

In particular, ranking the main causes for dissatisfaction shows that:
1. As stated before, computational scientists are most dissatisfied when their codes are not open source and offer slow performance.
2. Lack of active development appears as a notable second reason for dissatisfaction.
3. Errors that cannot be fixed was a major issue for many users.

In contrast, the following factors make for great nuclear code: commercial-grade dedication (compliance with regulatory needs for production), CAD input, and great UI/UX. The first, commercial-grade dedication, follows a need to cut-down time to real implementation of fission and fusion reactors and devices: a code that can be used in production is a code that will be favored by computational scientists. #2 and #3 were revelatory reasons for dissatisfaction, though. CAD input indicates a need for codes that can not only product simulations but visualize them according to CAD inputs. Great UI/UX indicates a deeper pain of computational scientists that merits more investigation, since very seldom do any of these codes have a visual UI, or UX designers employed for their development and improvement.

| Rank | Factor | Score (weighted by the position in each respondent's ranking) |
|---|---|---|
| 1 | Commercial grade dedication (e.g., NQA-1) | 322 |
| 2 | CAD input | 314 |
| 3 | Great UI/UX (User Interface/User Experience) | 296 |
| 4 | Multiple integrations with other codes | 279 |
| 5 | Simple extensibility (the ability to add new code or plugins) | 262 |
| 6 | Input is a standard programming language (e.g., Python) | 261 |
| 7 | Extensive validation | 240 |
| 8 | Easy to use APIs (Application Programming Interface) | 228 |
| 9 | Readable documentation | 225 |
| 10 | Other (please specify) | 214 |

*Table 2: Commercial-grade dedication and CAD input are the most desirable usability features a code can have.*

In the "Other" category, the following trends, clustered through the use of ChatGPT for codification, emerged:

| Multiphysics Integration | Open Source Development | AI and Machine Learning Integration |
|---|---|---|
| This trend is consistently mentioned as a crucial development. Paraphrasing, respondents are excited about:<br><br>● Coupling different physics models or codes<br>● Creating more comprehensive and realistic simulations<br>● Investigating how interacting mechanisms | This is by far the most frequently mentioned trend across the responses. Many respondents view open source as exciting because it:<br><br>● Allows for large community collaboration<br>● Eliminates barriers to access<br>● Enables easier extensibility and customization<br>● Consolidates intellectual efforts across different labs and | While not mentioned as frequently as Multiphysics Integration and Open Source Development, this trend is notable for its potential impact. Respondents are interested in:<br><br>● Accelerating expensive nuclear models<br>● Enhancing safety analysis<br>● Optimizing operations |
| affect quantities of interest<br>● Resolving locally occurring, complex phenomena | organizations | ● Developing surrogate models<br>● Potentially automating input generation |

## VI. DISCUSSION

The survey yielded important results that shine a light on different aspects of the computational nuclear science and engineering community. Let's begin with the composition of the respondents themselves. 80% of all computational scientists that responded to the survey, regardless of title or organization type, write code as a part of their jobs. In other words: it is possible that even managers in computational science are expected to continue programming well into their careers, a phenomenon that is highly debated in the broader software engineering community ([18]). This is compounded by the evidence that 45% of the respondents that contribute code do so to new code being created from scratch, and that regardless of the size of the company, novel code creation seems to be in a healthy state. Once again, though, we must look at these results as a reflection of the innovative nature of labs and universities, which compete for resources and recruits through their reputation for being at the bleeding edge of research and were the main employers of a majority of our respondents. If research labs require attracting top talent in computational science, they will probably do so by creating new codes that represent the state-of-the-art.

The above point is relevant because the software engineering industry, as a whole, tends to debate heavily on the need for engineering managers to program or not. [19] Part of the reason is that the industry has a broader divide, organizational and in terms of professional maturity, between coders (usually younger) and managers (older, tenured, expected to mentor and not create). There isn't consensus amongst the broader community on what an "ideal" manager is, whereas nuclear science managers seem to find themselves in the hybrid player/coach role.

The purported main advantage to such a mixed practice/management role is that it keeps managers close to the details of what they are coding [18] [20]. A bigger conversation, though, is whether such a short distance to the code creates better design decisions that are both empathetic of the final users (reactor designers, scientists, engineers, etc.) and beneficial for society as a whole, if, in the context of affordances of software, such empathy with a human user will lead to more empathetic, better designer nuclear science.

Another point for reflection is the fact that, while the majority of respondents claimed that more that more than 10 people in their organization tend to use the codes they create in their organization (Fig.7), 24.3% do not have a large user base. Considering the vast majority of our respondents work in companies with over 500 people, this could indicate that a) organizations are not heavily investing in software or b) that a single organization might have many small teams working with many disparate codes, inside the most innovative organizations. Though we (the researchers) have heard of the latter as a problem in the field of computational science, specifically about the competitive nature of teams even inside a single organization, more research should be made on this point in future semi-structured interviews.

Moving on to the problems being solved by respondents as a part of their jobs, it is important to note that 7 respondents mentioned working "all of the categories" of inquiry (Fig. 8). This means that some computational scientists are creating codes that tackle parametric study for design, parametric study for numerical methods, safety analysis, and uncertainty analysis. 2 more respondents were specifically focusing on multiphysics and multiscale modeling, indicating the rise of multi-analysis code as an interesting trend that will probably become the norm as codes being currently developed in universities and national labs become adopted more broadly.

Separately, this trend toward multi-analysis codes continues in the rest of our results. More and more scientists are focused on developing multiphysics code. This trend of multiphysics code creation dovetails nicely into the topic of code choice: the rise in use of OpenMC and BISON matches an overall shift towards multiphysics code adoption. An exciting one too, in our opinion, since it matches the current proclivity in the broader software technology industry to making things smaller, modular, and more general-purpose.

Certain results did cause surprise, though: seeing Python and C++ as the most utlized programming language contrasted with the informal interviews with experts in computational scientists that we had while developing the questions for our survey; we expected Fortran to take a spot in top two answers.. It's possible that the exclusion of Fortran from the top arises from the bias of our respondent profile (university and national lab computational scientists), and matches the overall trend on developing or using multiphysics codes in labs and universities. But, our initial informal interviews with experts to understand the field of computational science were clear: interviewees expected Fortran to dominate codes, while expressing dismay at the outdatedness of the language, which is 68 years old. The fact that labs and universities, which are usually at the forefront of innovation in fusion and fission, are favoring Python and C++, is a heartening look into the future: perhaps there is an upcoming renaissance of programming languages applied to fusion and fission.

An exciting discovery from this survey was the topic of budget allocation to computational science. The median response for the estimated budget spent on nuclear engineering software is $1 million dollars, which is already significant considering that the average software developer tends to make around $200k a year (and nuclear computational science salaries tend to be around two thirds of that). Conservatively, we are looking at teams of around 3-5 people in each organization developing codes. Of course, there are exorbitant budgets for bigger organizations (the largest results, directly from our survey, being fifty million dollars!), but even the mean budget of $1 million dollars is significant in a world where startups in nuclear are being founded and invested in at an accelerated pace [21].

Finally, delving into the reasons for satisfaction (or dissatisfaction) with codes yielded a couple of surprising results. When asked what their top 3 reasons for being dissatisfied with their code of choice is, respondents suggest the lack of open-source access, slow performance, limited development activity, and unresolved errors are significant concerns for users. In our original talks with experts, to help guide the design of this study, they mentioned they suspected open-source access would not be a priority. Conversely, only one of the 3 experts we chatted with expected slow performance to be a major sticking point for computational scientists. It might be that open-source is seen as a driving factor for speed, since scientists would be able to customize open-source code to their liking. But more interviews should be conducted to explore whether or not this is a true correlation.

On the other hand, the reasons for appreciating codes were surprising to the PI: the first, commercial grade dedication, is self-explanatory: a code that can be quickly used in production is a code that will be favored by computational scientists. #2 and #3 were surprises, though. CAD input indicates a need for codes that can not only product simulations but visualize them according to CAD inputs. Great UI/UX indicates a deeper pain of computational scientists that merits more investigation, since very seldom do any of these codes have a visual UI, or UX designers employed for their development and improvement.

Overall, more than a snapshot of the present, the results of the survey represent a look into what will happen 5 to 10 years in the nuclear industry, and, more broadly, the future energy landscape of the world. Since the majority of our respondents belonged to US national labs and universities, these results hint at the most cutting-edge trends in the industry.

## VII. LIMITATIONS

One clear limitation of this study is the total number of people surveyed. Though 103 respondents is a statistically non-representative stratified sample, ideally we would want to increase this number every year, and in particular include more voices from energy providers instead of almost exclusively academia and research laboratories.

Additionally, the majority work in fission, not fusion. This risks leaving out important insights into how computational science is developed in that nascent industry.

One admitted flaw of our survey design is that plasma physics could have been better included in our possible answers. We consider this to be a minor issue, though, since across scientists working on both fusion and fission as a part of their jobs, multiphysics continues to be a priority, regardless of the technical differences of multiphysics codes for fusion and fission.

Finally, all of our respondents work in US-based organizations. A true survey of the computational science industry in fusion and fission can, and should, include a more international perspective.

## VIII. CONCLUSION

As a broader conclusion, it is highly possible that computational science determines the future affordances of clean energy such as fusion and fission. Understanding the innovative nature of new codes, and how they are produced in labs and universities, gives this idea further strength. Considering that there is a rapid rise in job opportunities for scientists working with codes incubated in labs and academia appears to back up this inference.

However, not enough money has been spent on the topic of software tools utilized by fusion and fission organizations, as indicated by this being the first survey in the area. This means that any findings in this survey might need to be further confirmed with qualitative data and future surveys.

With those caveats in mind, the fact that the third most important reason to choose a code is great UI/UX, and the fifth one is simple extensibility, might indicate the importance of basic DevEx of codes, and perhaps even a call to make them far more user-friendly. To solve this question, the future might include more research into whether current codes feel UI/UX appropriate or not, and what that means specifically for computational scientists.

Additionally, there seems to be an appetite for codes that help design smaller devices and reactors, as is made clear by the overall top scores BISON and OpenMC get for desirability to work them during 2024. When putting this side by side with the fact that many respondents asked to focus on those two particular codes, it might indicate that either BISON and OpenMC have good DevEx, or that since they are used more often, the DevEx of those two codes is slowly shaping the perception of what "good" and "bad" codes are and can be.

One place where our study is lacking is the respondent pool by organization. Since no computational scientists working in energy providers responded to the survey, we don't have a full picture of the present with these survey results, only of what researchers are making strides in, and perhaps will release into the public in a couple of years. Regardless, this survey represents the first-ever breakdown of the demographic characteristics of computational scientists working in fusion and fission, so in spite of the lack of energy provider scientists in the pool of respondents, the survey is a clear advancement towards establishing the research topic of fusion and fission computational science.

Finally, this survey uncovered a look at the main trends to come in this 2025: the rise of open-source software, the rapid adoption of multiphysics software, and the coupling of AI with fusion and fission codes. We are looking forward to this year's survey to find out if the trends have become a reality by end of year.

## VIII. FUTURE WORK

In the immediate future, the team wishes to perform two new pieces of research:
1. Semi-structured interviews with the respondents that agreed to be interviewed in the last question of our survey.
2. A follow-up survey looking back at 2024 and forward to the rest of 2025. This survey would continue to be repeated every year.

### VIII.A Semi-Structured Interviews

25 respondents expressed their willingness to being interviewed as a follow-up to the survey. A couple of thematics to delve into more deeply on a qualitative level are:

| Topic | Potential Questions |
|---|---|
| Most "controversial" codes (codes that generated the biggest split in opinion) | - What is lacking, in your opinion, in these codes: Griffin, MCNP, and OpenMC.<br>- On the flipside, what is successful in those codes?<br>- If you had a magic wand, what would you change about each of those codes? |
| UX/UI likes and dislikes | - Please prioritize these specific aspects of UI/UX in codes, from 1 to 6: accessibility, color palette, layout, multi-screen capability, UX flow. |

|                                | - Here are pictures of each code in each aspect. Have you ever used this code in this screen? What's your opinion on it?<br>- What is the thing you dislike the most, from a UX/UI standpoint, about this code?<br>- If you could draw a UI for this code, what would it look like? Who would you make it for? |
|--------------------------------|---|
| Code usage inside organizations | - How many people use the code you've created inside your organization?<br>- What does each of them do?<br>- How do they interact with your code?<br>- Have they expressed what they like or don't like about the code?<br>- Are there competing code-development teams inside the organization? Or is your team the only team doing so? |
| Development pain points        | - What is one aspect of the code you create that most users would not know about?<br>- Is there a design decision you made while creating the code that you regret? Why? What would you have done differently?<br>- Who do you imagine your ideal user to be? And then, what about your "real" user? |
| Open-source                    | - Why do you use open-source codes? In your opinion, what are their advantages? What are their disadvantages<br>- Do you want to use more or less open-source codes? |

### VIII.B. Annual Surveys

Every year, a new survey with the exact same questions will be sent out to the broader computational science community in fusion and fission. The objective will be to gather more information from two particular audiences: energy providers, and students (since 70% of all respondents for 2023-2024's survey were not students). We hope this will continue to refine our results. Additionally, new survey questions might arise from the semi-structured interviews, and we are open to such contributions.

ering-managers-is-he-right/; (current as of Nov. 22, 2024).
21. "A new era for nuclear energy beckons as projects, policies and investments increase," IEA; https://www.iea.org/news/a-new-era-for-nuclear-energy-beckons-as-projects-policies-and-investments-increase; (current as of Apr. 15, 2025).

## IX. ACKNOWLEDGMENTS

The research team would like to thank, in no particular order: Derek Gaston, April Novak, Patrick Shriwise, for answering informal questions as the original survey was designed.

## X. APPENDIX

### X.I. APPENDIX A: Glossary of programming languages

| Language | Definition |
|---|---|
| C | A low-level, procedural programming language known for its efficiency and close-to-hardware memory management. Widely used in high-performance computing (HPC) and embedded systems, including legacy nuclear simulation codes. |
| C++ | An extension of C that supports object-oriented programming. It combines fine-grained memory control with high-level abstractions, making it suitable for large-scale simulation frameworks and computational physics applications. |
| Fortran | A high-performance language originally developed for scientific and numerical computation. Still heavily used in nuclear engineering and physics due to its optimized handling of array operations, numerical solvers, and legacy codebases (e.g., MCNP, SERPENT). |
| Julia | A high-level, high-performance language designed for numerical and scientific computing. It offers just-in-time (JIT) compilation via LLVM and combines the speed of C with the expressiveness of Python, making it increasingly relevant for modern nuclear |

|   | data analysis and modeling. |
|---|---|
| MATLAB | A proprietary language and environment optimized for matrix computations and algorithm prototyping. Often used for control system modeling, signal processing, and reactor dynamics studies, though less common in large-scale HPC contexts due to performance limitations. |
| Python | An interpreted, high-level language widely adopted for scripting, automation, data analysis, and machine learning in scientific research. While not optimized for raw performance, it interfaces well with C/C++ and Fortran libraries via tools like NumPy, SciPy, and f2py, making it valuable for prototyping and post-processing. |

## X.II APPENDIX B: Survey design questions in full

**Q0.**

In this survey, we (Andrea Morales and Aditi Verma, co-investigators, as part of Fastest Path to Zero) aim to determine trends in usage of fission and fusion software. No similar survey has been created before, and we believe setting the stage for how code is used in fusion and fission will help create more user-friendly codes in the future.

For clarification, in the survey we refer to "fission or fusion code" interchangeably with "fission or fusion software" to mean "a program of software used to solve or analyze a problem in the fusion or fission engineering and design process."

If you agree to participate in this survey, you will be asked to answer questions around your usage of software in your work setting. Your answers will not be compensated. Participating in this study is completely voluntary. Even if you decide to participate now, you may change your mind and stop at any time.

This survey is completely anonymous and does not require any identifiable personal information. In its final question, you can opt into providing your contact information for further in-person interviewing, if you so wish, but it is not obligatory for you to do so. If you do opt in, your information will only be utilized to get in touch with you for further research, and will not be disclosed to any third-parties.

Information collected in this project may be shared with other researchers, but we will not share any information that could identify you. If you have questions about this research study, please contact Andrea Morales Coto at amcoto@umich.edu. As part of their review, the University of Michigan Institutional Review Board Health Sciences and Behavioral Sciences has determined that this study is no more than minimal risk and exempt from on-going IRB oversight.
- I understand, and wish to start the survey
- I decline participation

**About You**

**Q1.**

What field of energy do you work in?
- Fission
- Fusion
- Both
- Other (please specify)

**Q2.**

What is your area of specialization?
- Device Design
- Reactor Design
- Materials Engineering
- Computational Science

- Thermohydraulics
- Plasma Physics
- Other (please specify)

**Q3.**

What is the nature of your involvement with fission or fusion software creation?
- I have programmed new software from scratch.
- I have contributed code to software that is already existing.
- I have not programmed from scratch or contributed to software.

**Q4.**

Do you use fusion or fission software at work?
- Yes, as a daily part of my job.
- I used to, but not anymore.
- No, and I never have.

**Q5.**

Have you programmed software for fission or fusion that is currently used by more than 10 people?
- Yes, my code is used by more than 10 people.
- No, my code is used by less than 10 people.

**The Problem You're Solving with Code**

**Q6.**

Please describe the problem you are trying to solve by using software in your field.

**Q7.**

What type of problem/analysis are you doing?
- Safety analysis
- Uncertainty analysis
- Parametric study for numerical methods Parametric study for design
- Other (please specify)

**Q8.**

What physics of the system are you mostly focused on?
- Fuel performance
- Thermal-hydraulics, thermal-fluids, heat transfer System level analysis
- Depletion/transmutation
- Multiphysics
- Materials performance
- Neutronics

**Q9.**
How long have you been involved in your professional field (including academic studies)?
- Less than a year
- Between a year and 2 years
- >2-5 years
- >5-10 years
- Over 10 years

**Your Organization**

**Q10.**
How long have you been involved in your professional field (including academic studies)?
- Less than a year
- Between a year and 2 years
- >2-5 years
- >5-10 years
- Over 10 years

**Q11.**
Are you currently enrolled in an academic institution as a student, in a field directly related to fusion or fission?
- Yes, I am currently a student.
- No, I am not a student.

**Q12.**
Are you currently working professionally (not as a part of your studies)?
- Yes, I am working and studying.
- No, I am exclusively studying.

**Q13.**
What type of organization are you currently a part of?
- Startup
- National Lab
- University
- Think Tank
- Energy Provider
- Other (please specify)

**Q14.**
What is the size of the organization you are a part of?
- 1-10 employees
- 11-50 employees
- 51-100 employees

- 101-500 employees
- 501+ employees

**Your Software Tools**

**Q15.**
What are the main programming languages you use at work?
- Python
- Fortran
- C++
- Julia
- MATLAB
- C
- None - I don't program
- Other (please specify)

**Q16.**
During a normal week of work, what software do you tend to use?
- My employer's code
- NRC codes: SCALE/Polaris, PARCS, Bluecrab, TRACE
- VERA, MPACT, CTF
- NEAMS Codes: SAM, Griffin, Sockeye, BISON, PRONGHORN
- SCALE
- MCNP
- Open Source: Cyclus, OpenMC, OpenMOC, Gen-FOAM
- Vendor Independent Commercial Code (e.g., CASMO, SIMULATE, ANSYS) International Code: Serpent, Dragon, nTRACER
- Legacy codes: DIF3D, SAS, RELAP
- Other Research codes: COMET, PENTRAN, NEM, RAPID

**Q17.**
Please specify how often during the week you use each of the codes you selected in the previous question. This question automatically populates with the options you selected before.

|  | **Never Used** | **Sometimes Used** | **Always Used** |
| --- | --- | --- | --- |
| MCNP |  |  |  |
| SCALE |  |  |  |
| VERA MPACT |  |  |  |
| CTF |  |  |  |
| My employer's code [Insert your own] | | | |
| **NRC Codes** | | | |
| SCALE/Polaris | | | |
| PARCS | | | |
| Bluecrab | | | |
| TRACE | | | |
| NEAMS Codes | | | |
| SAM | | | |
| Griffin | | | |
| Sockeye | | | |
| BISON | | | |
| PRONGHORN | | | |
| **Open Source** | | | |
| OpenMOC | | | |
| Cyclus | | | |
| OpenMC | | | |
| Gen-FOAM | | | |
| **Vendor Independent Commercial Code** | | | |
| CASMO | | | |
| SIMULATE | | | |
| ANSYS | | | |
| **International Code** | | | |
| Serpent | | | |
| Dragon | | | |
| nTRACER | | | |
| **Legacy Codes** | | | |
| SAS | | | |
| RELAP | | | |
| DIF3D | | | |
| **Other Research Codes** | | | |
| COMET | | | |
| PENTRAN | | | |
| NEM | | | |
| RAPID | | | |

## Q18.

Thinking about the following programming languages, which ones did you work with last year, and which ones would you desire to work with the rest of this year?

| | Worked with in 2023 | Want to work with in 2024 |
|---|---|---|
| MATLAB | | |
| C | | |
| Julia | | |
| Python | | |
| C++ | | |
| Fortran | | |

**Q19.**

Now, thinking about fusion and fission software, which codes did you work with last year, and which ones would you desire to work with next year?

|  | Worked with in 2023 | Want to work with in 2024 |
|---|---|---|
| MCNP |  |  |
| SCALE |  |  |
| VERA |  |  |
| MPACT |  |  |
| CTF |  |  |
| Insert your own |  |  |
| SCALE/Polaris |  |  |
| PARCS |  |  |
| Bluecrab |  |  |
| TRACE |  |  |
| NEAMS Codes |  |  |
| SAM |  |  |
| Griffin |  |  |
| Sockeye |  |  |
| BISON |  |  |
| PRONGHORN |  |  |
| OpenMOC |  |  |
| Cyclus |  |  |
| OpenMC |  |  |
| Gen-FOAM |  |  |
| CASMO |  |  |
| SIMULATE |  |  |
| ANSYS | | |
| Serpent | | |
| Dragon | | |
| nTRACER | | |
| SAS | | |
| RELAP | | |
| DIF3D | | |
| COMET | | |
| PENTRAN | | |
| NEM | | |
| RAPID | | |

**Focus on a single code**

**Q20.**
 In an earlier question, you mentioned you use certain codes most often. Please choose one to focus on for this section.
- MCNP
- SCALE
- VERA
- MPACT
- CTF
- Insert your own
- SCALE/Polaris
- PARCS
- Bluecrab
- TRACE
- SAM
- Griffin
- Sockeye
- BISON
- PRONGHORN
- OpenMOC
- Cyclus
- OpenMC
- Gen-FOAM
- CASMO
- SIMULATE

- ANSYS
- Serpent
- Dragon
- nTRACER
- SAS
- RELAP
- DIF3D
- COMET
- PENTRAN
- NEM
- RAPID

**Q21.**

Were you involved in the original process of creating that code?
- Yes
- No

**Q22.**

Who made the decision in your organization to use that code?
- Heads of the organization (executives, or heads of department)
- Engineering/computational science team lead
- Myself
- Other (please specify)

**Q23.**

What are your top 3 reasons for *not* being satisfied with the software you use most often?
- Reason 1 (causing the most dissatisfaction):
- Reason 2:
- Reason 3:

**What makes great code**

**Q24.**

In your opinion, which of the following factors make for great nuclear code? Rank the following options, #1 being the most influential factor for a code to be considered "great", and the last being the least influential factor.
- Great UI/UX (User Interface/User Experience)
- Easy to use APIs (Application Programming Interface)
- Readable documentation
- Multiple integrations with other codes
- Extensive validation
- Simple extensibility (the ability to add new code or plugins)
- Commercial grade dedication (e.g., NQA-1)
- CAD input
- Input is a standard programming language (e.g., Python)

- Other (please specify)

## Q25.

In your opinion, what is currently the most exciting trend in software for nuclear engineering? Why?

## Q26.

Last question! We want to continue to do research in this topic, and make software for nuclear engineering more user friendly. Would you like to be contacted in the future for further interviews on the topic of nuclear codes?
- Yes, please contact me for further interviews
- No, please do not contact me

# APPENDIX C: All reasons for user dissatisfaction with codes

| Reason 1 for Dissatisfaction | Reason 2 for Dissatisfaction | Reason 3 for Dissatisfaction |
|---|---|---|
| Continuous development and lack of documentation may raise unexpected bugs and user confusion | Lack of user support | Lack of examples |
| Legacy interface and coding | N/A | N/A |
| difficulty in getting access | undocumented features | system portability |
| Mesh generation workflows are cumbersome | Lack of good documentation | Lack of developer support |
| Not fit for deep penetration problems | Bugs with unstructured mesh MPI scaling | Low ability to tailor source descriptions in mesh geometry on an element by element basis |
| Poorly written | Unmaintained | Undocumented |
| Not all software is open source. | Many of them are dificult to get installed. | Not interested to buy, I am a poor guy. |
| Consistency of API | Performance | Installation process |
| Not enough resources for user support | Complex code, hard to modify easily | None- quite happy overall |
| not open source | N/A | N/A |
| Very happy with our software! | NA | NA |
| slow | not user friendly | not validated |
| not converging | developments needed, no funds for dev | no input validation |
| Installation/distribution challenges | Documentation challenges | Difficulty in maintaining OS software |
| Im actually satisfied with openmc and if im not then i will improve it till im satisfied | Still satisfied | Yep also still satisfied |
| Development funding challenges | Licensing headaches | Validation maintenance |
| runtime | memory usage | ease of use |
| Closed source. Source access requires review. Limits collaboration. | N/A | N/A |
| not open source | input format is antiquated | output format is also antiquated |
| NA | NA | NA |
| More funding needed | - | - |
| N/A | N/A | N/A |
| Learning curve | Expert knowledge required for settings | N/A |
| NA | NA | NA |
| Convergence issues with multiphysics nonlinear solve | Small timesteps due to poor conditioning of the nonlinear solve. | Ease of setting up new problems from scratch without a GUI. |
| Bugs sometimes occur | Lack of features pertaining to my application | N/A |
| Extremely antiquated | Closed-source | No native API |
| Some lack of error checking in python API | lack of charged particle support | None: I love OpenMC |
| Computational cost | Scalability | Fidelity |
| Formatting/naming convention updates | Slow merging of new features | Lack of fusion specific examples |
| Continuous change of version | Funds to fix the code since it is composed by many sub-modules | Lack of clear and stable manual |
| n/a | n/a | n/a |
| Not being able to use one cell in multiple universes. (kinda overwrites the code and forgets the first universe!) | A little lack of guidance or tutorial. | Updating openmc when it is downloaded from source. |
| n/a | n/a | n/a |
| slow | not gpu accelerated | memory consumption |
| convergence | stabiilty | mechanical capabilities |
| There are not enough built-in functions which are necessary for calculating parameters | N | N |
| Vague error messages | Difficulty in coupling | Sparse comments |
| Long time to compile | Git submodule update process can be cumbersome | Lack of GUI |
| Still limited documentation | Time required for meshing | Still limited GPU support |
| Requires to know how to program in C++ | Takes a while to learn how to use it | There is no user interface |
| Input complexity | . | . |
| OpenFOAM structure | OpenFOAM code support | OpenFOAM duality |
| Lack of open source | Lack of active development | Errors that cannot be fixed |